\documentstyle[epsfig]{mn}

\title{A {\sc{robust}} method for fitting peculiar velocity field models}
\author[S. Rauzy and M. A. Hendry]
       {St\'ephane Rauzy and Martin A. Hendry\\
        Department of Physics and Astronomy, University of Glasgow,
Glasgow, G12 8QQ, UK.}
 \date{Accepted .
      Received ;
      in original form }

\begin{document}

\maketitle

\begin{abstract}
    
We present a new method for fitting peculiar velocity models
to complete flux limited magnitude-redshifts catalogues, using
the luminosity function of the sources as a distance indicator.
The method is characterised by its robustness.
In particular, no assumptions are made concerning
the spatial distribution of sources and their luminosity
function. Moreover, selection effects in redshift are allowed.
Furthermore the inclusion of additional observables correlated with
the absolute magnitude --
such as for example rotation velocity information as described by the 
Tully-Fisher relation -- is straightforward.   

As an illustration of the method,
the predicted IRAS peculiar velocity model characterised by the density
parameter $\beta$ is tested on two samples.
The application of our method to the Tully-Fisher MarkIII MAT sample leads
to a value of $\beta=0.6 \pm 0.125$, fully consistent with the results
obtained previously by the VELMOD and ITF methods on similar datasets. 
Unlike these methods however, we make a very conservative use of the 
Tully-Fisher information. Specifically, we require to assume neither
the linearity of the Tully-Fisher relation nor a gaussian distribution
of its residuals. Moreover, the robustness of the method implies that no
Malmquist corrections are required.

A second application is carried out, using the fluxes of the IRAS 1.2 Jy 
sample as the distance indicator. In this case the effective depth of the 
volume in which the velocity model is compared to the data is almost twice 
the effective depth of the MarkIII MAT sample. The results suggest that the 
predicted IRAS velocity model, while successful in reproducing locally the 
cosmic flow, fails to describe the kinematics on larger scales.

\end{abstract}

\begin{keywords}
cosmology: large-scale structure of Universe -- 
galaxies: distances and redshifts -- 
methods: statistical, data analysis
 \end{keywords}

\section{Introduction}
\label{Introduction}

The study of the large-scale motions of galaxies in the Universe
may provide valuable information concerning the dynamics of large-scale
structures and the nature of the underlying dark matter. 
According to the gravitational instability scenario, the peculiar velocity
field (i.e. the deviation from the smooth Hubble flow) may be used
to infer the power spectrum of the mass fluctuations on intermediate scales
and to constrain the cosmological density parameter $\Omega$
(see for example Dekel 1994).

Since the discovery of the Great Attractor (Lynden-Bell et al. 1988),
the field has proven to be particularly active.

Various observationnal programs have been completed, providing large
and accurate datasets: e.g. the W91CL and W91PP samples (Willick 1990);
MAT sample (Mathewson et al. 1992); 
HM sample (Han \& Mould 1992); 
CF sample (Courteau et al. 1993); 
Abell BCG sample (Lauer \& Postman 1994); 
SCI sample (Giovanelli et al. 1997a);
KLUN sample (Theureau et al. 1997); 
nearby SNIa sample (Riess et al. 1997);
MarkIII dataset (Willick et al. 1997b); 
SBF survey (Tonry et al. 1997);
SFI sample (Giovanelli et al. 1998); 
SMAC sample (Hudson et al. 1999);
EFAR and ENEAR samples (Colless et al. 1999 and Wegner et al. 1999);
SCII sample (Dale et al. 1999); 
Shellflow survey (Courteau et al. 1999);
LP10k survey (Willick 1999a).

Another significant advance during the past decade has been the improved
understanding of the statistical formalism underlying the use of galaxy
distance indicators -- and in particular the principles and practical methods
of correcting for Malmquist bias. (See for example
Hendry \& Simmons 1990,
Teerikorpi 1990,
Bicknell 1992,
Landy \& Szalay 1992,
Triay et al. 1994,
Willick 1994,
Hendry \& Simmons 1994,
Sandage 1994,
Willick et al. 1995,
Freudling et al. 1995,
Willick et al. 1996,
Rauzy \& Triay 1996,
Ekholm 1996,
Triay et al. 1996,
Rauzy 1997,
Willick et al. 1997b,
Giovanelli et al. 1997b,
Theureau et al. 1998,
Teerikorpi et al. 1999).

Several methods for extracting dynamical and kinematical information
from distance indicator datasets have been proposed, i.e.
the POTENT method (Bertschinger \& Dekel 1989,
Dekel et al. 1990,
Bertschinger et al. 1990,
Dekel et al. 1999 and references therein) 
and its variants 
(Rauzy et al. 1993 and 1995, Newsam et al. 1995); 
the ITF method (Nusser \& Davis 1995,
Davis et al. 1996, Da Costa et al. 1998); 
the VELMOD method (Willick et al. 1997a, Willick \& Strauss 1998).

The comparison between the peculiar velocity or density fields
inferred from distance indicator data with their corresponding
fields derived from whole-sky redshift surveys has been one
of the major issues addressed throughout the last decade.
The question here is whether the spatial distribution of luminous
matter e.g. the galaxies,
traces the underlying mass fluctuations and if not, what are
the properties of the ``biasing'' between the two fields?

Up to now, the point has not received any consensual answer.
Indeed, the application of POTENT to various distance indicator
datasets (Sigad et al. 1998 and references therein) favours
a value of $\beta_I = \Omega_0^{0.6}/b_I \simeq 1$ for the linear ``biasing''
density parameter, while the VELMOD and ITF fitting methods
lead to a value of   
$\beta_I \simeq 0.5$ 
(Davis et al. 1996, 
Willick et al. 1997a, Riess et al. 1997,
Da Costa et al. 1998, 
Willick \& Strauss 1998). The origins of this 
significant
discrepancy have not yet been elucidated (see for example
Strauss 1999, Willick 1999b).
At least one of these methods is suffering from 
some systematic effects, i.e. some statistical biases plaguing the 
estimate of the parameter $\beta$ and
not accounted for in the error analysis.

This remark leads us to the object of the present paper. 
The POTENT, VELMOD and ITF methods all require, at some
stage of the analysis, to assume some a priori working hypotheses concerning
the characteristics of the distance indicator dataset.
For Tully-Fisher data for example, it will be assumed 
that the Tully-Fisher law is well described by a linear relation.
These methods apply moreover
under the hypothesis that the observational selection effects
obey particular conditions. 
How are the results affected if one or more of the working assumptions
fails to be satisfied by the dataset is generally a question
not addressed in the error analysis.

The philosophy of the method we present herein is to reduce as far
as possible the number of a priori hypotheses concerning the
distance indicator sample. A direct consequence is that
the range of application of the method will be considerably
broadened.

The statistical background of the method is presented section 2.
Its potential is illustrated by testing
the predicted IRAS peculiar velocity model
on two samples.
In section 3, we perform the analysis  
using the fluxes of the IRAS 1.2 Jy survey as the distance indicator.
We have deliberately chosen this sample looking to demonstrate
the wide range of application of the method. Where the POTENT, VELMOD
or ITF methods would not have been successful in extracting 
kinematical information 
from this dataset, our method does.
We also treat a more classical case, 
the Tully-Fisher MarkIII MAT sample, in section 4. Finally, in section 5 we
summarise our conclusions.

\section{The Method}

\subsection{The statistical model}

It is assumed herein that the
distribution function of the absolute magnitudes $M$
of the
population, i.e. the luminosity function $f(M)$, does not depend on the
spatial position ${\bf r}=(r,l,b)$ of the galaxies. The probability density
describing the sample splits in this case as
\begin{equation}\label{dP_1}
dP \propto dP_{\bf r} \times dP_M= \rho(r,l,b)\,r^2 \cos b \,
dl db dr \times
f(M) dM
\end{equation}
where $\rho(r,l,b)$ is the spatial distribution function of the sources.

The application of the method will be restricted to
samples strictly complete up to
a given magnitude limit $m_{\rm lim}$, or in other words where the selection
function in apparent magnitude is well described by a sharp cut-off, i.e.
$\psi(m)=\theta(m_{\rm lim}-m)$ with
$\theta(x)$ the Heaviside function. Accounting for selection
effects, the probability density of the sample may be rewritten as
\begin{equation}\label{dP_2}
dP = \frac{1}{A} h(\mu,l,b) \cos b \,dl db d\mu \, f(M) dM
\, \theta(m_{\rm lim}-m)
\end{equation}
where $\mu=m-M =
5\,\log_{10} r + 25$ is the distance modulus, $h(\mu,l,b)$ the line-of-sight
distribution of $\mu$ and $A$ is the normalisation
factor warranting $\int dP =1$. For convenience in notation
the angular dependence in $l$ and $b$ will be hereafter implicit.
Observational selection effects in apparent magnitude then introduce
a correlation between $M$ and $\mu$.

The milestone of the method consists in defining  the random
variable $\zeta$ as follows
\begin{equation}\label{zeta}
\zeta = \frac{F(M)}{F(M_{\rm lim})}
\end{equation}
where $F(M)=\int_{-\infty}^M f(x)dx$ stands for the cumulative
distribution function in $M$ and $M_{\rm lim}\equiv
M_{\rm lim}(\mu)$ is the maximum absolute magnitude for which
a galaxy at distance $\mu$ would be visible in the sample
(e.g.  $M_{\rm lim}(\mu)=m_{\rm lim} - \mu$ if the k-correction is
neglected). The volume element may be rewritten as
\begin{equation}\label{dzetadmu}
d\mu \,d\zeta = \frac{f(M)}{F(M_{\rm lim}(\mu))} \,d\mu \,dM
\end{equation}
and by definition the random variable $\zeta$ for a
sampled galaxy belongs to the interval $[0,1]$. The probability
density of Eq. (\ref{dP_2}) reduces thus to
\begin{equation}\label{dP_3}
dP =  \frac{1}{A} h(\mu)\,F(M_{\rm lim}(\mu))
\,d\mu \, \times \,
\theta(\zeta)
\, \theta(1-\zeta)
\, d\zeta
\end{equation}
with $A= \int h(\mu)\,F(M_{\rm lim}(\mu)) \,d\mu$. Note that the
probability density $
dP_\mu =  \frac{1}{A} h(\mu)\,F(M_{\rm lim}(\mu))
\,d\mu
$ describes the observed spatial distribution function of the sources.
It follows from Eq. (\ref{dP_3}) that:
\begin{itemize}
\item P1: $\zeta$ is uniformly distributed between $0$ and $1$.
\item P2: $\zeta$ and $\mu$ are statistically independent,
i.e. the distribution of $\zeta$ does not depend on
the spatial position of the galaxies.
\end{itemize}
Property P1 may be used to construct a test for assessing the
completeness of the sample in apparent magnitude. The details of this test
are presented in a separate paper (Rauzy, in preparation), although we apply
it to the Mark III MAT sample later in this paper.  
The new method we propose hereafter for fitting peculiar velocity field models 
is based on
property P2.

\subsection{Estimate of the random variable $\zeta$}

The random variable $\zeta$ can be estimated without any prior
knowledge of the cumulative luminosity function $F(M)$. To 
each data point with coordinates $(M_i,\mu_i)$ is associated the region 
$S_i=S_1 \cup S_2$ defined as 
\begin{itemize}
\item $S_1 =  \{\,(M,\mu) {\rm ~such~that~}
 M \le M_i {\rm ~and~} \mu \le \mu_i \,\}$
\item $S_2 =  \{\,(M,\mu) {\rm ~such~that~} M_i <  M \le M^i_{\rm lim}
 {\rm ~and~} \mu \le \mu_i \,\}$
\end{itemize}
The random variables $M$ and $\mu$ are independent in each subsample $S_i$ 
since by construction the cut-off in apparent magnitude is superseded
by the constraints 
$ M \le M^i_{\rm lim}$ and $\mu \le \mu_i$
(see figure 1). 
This implies that the number of points $r_i$ belonging to $S_1$ is 
proportional to $\int_{-\infty}^{M_i} f(M)\,dM = F(M_i)$, the number of
points $n_i$ in $S_i=S_1 \cup S_2$ is proportional to
$F(M_{\rm lim}^i)$ and that the quantity 
\begin{equation}\label{zetaestimate}
{\hat \zeta_i} =  \frac{r_i}{n_i+1}
\end{equation}
is an unbiased estimate of the
random variable $\zeta$. Equivalently  
the estimator ${\hat \zeta_i}$ may be defined as the normalised
rank of  
the point $M_i$ when the $M$'s are sorted
by increasing order within the subsample $S_i$
(see Efron \& Petrosian 1992).

\subsection{Radial peculiar velocity field models}

Let us first assume that the true radial peculiar velocity
field 
$v({\bf r})$ can be described by a one-parameter velocity model 
$ v_{\bf \beta}({\bf r})$, i.e. there exists a solution 
${\bf \beta}^{\star}$ satisfying 
$ v_{{\bf \beta}^{\star}}({\bf r}) \equiv 
v({\bf r})$.

For a given value of the parameter ${\bf \beta}$, the model dependent variables
$\mu_{\bf \beta}$ and
$M_{\bf \beta}$ can be computed (modulo the value of the Hubble
constant $H_0$) from the observed redshift $z$ and
apparent magnitude $m$ following
\begin{equation}\label{mubeta}
\mu_{\bf \beta}= 5\,\log_{10} \frac{cz}{H_0} + 25 -
u_{\bf \beta}
\,\,\,\,\,\,\,\,;
\,\,\,\,\,\,\,\,
M_{\bf \beta}= m - \mu_{\bf \beta}
\end{equation}
where the quantity
$u_{\bf \beta}$ is defined as
\begin{equation}\label{ubeta}
u_{\bf \beta}= - 5\,\log_{10} \left (1 - \frac{v_{\bf \beta} }{cz}
 \right )
\end{equation}
The quantities
$\mu_{\bf \beta}$ and
$M_{\bf \beta}$ are related to the true absolute magnitude $M$ and distance
modulus $\mu$ via
\begin{equation}\label{mubetabis}
\mu_{\bf \beta} = \mu  +
u_{\bf \beta^{\star}} -
u_{\bf \beta}
\,\,\,\,\,\,\,\,;
\,\,\,\,\,\,\,\,
M_{\bf \beta}
= M
- u_{\bf \beta^{\star}} +
u_{\bf \beta}
\end{equation}
Computing $\zeta_\beta$ from
$\mu_{\bf \beta}$ and
$M_{\bf \beta}$
as proposed in Eq. (\ref{zeta}) gives for the probability
density of Eq. (\ref{dP_3}),
\begin{equation}\label{dP_4}
dP =  \frac{1}{A} h(\mu)F(M_{\rm lim}(\mu_\beta))
\,d\mu \, \, C_\beta \,\, 
\theta(\zeta_\beta)
\theta(1-\zeta_\beta)
d\zeta_\beta
\end{equation}
where $C_\beta$ takes the following form when
$(u_{\beta^\star} 
-u_{\beta}) \ll 1$ (or equivalently
$(v_{\beta^\star} 
-v_{\beta}) \ll cz$),
\begin{equation}\label{Correlation1}
C_\beta = \frac{f(M)}{f(M_\beta)}
\simeq 1 +
(u_\beta
-u_{\beta^{\star}} )\,
\,\left (\ln f \right )^\prime (M_\beta)
\end{equation}
Because the absolute magnitude $M_\beta$, and hence the quantity
$\left (\ln f \right )^\prime (M_\beta)$,
is correlated with
the random variable $\zeta_\beta$,
$C_\beta$ acts as a correlation
coefficient between
$\zeta_\beta$  and the proposed velocity field model
$u_\beta$ when $\beta  \ne {\beta^{\star}}$. On the
other hand if
$\beta  = {\beta^{\star}}$ these quantities are statistically
independent, since in this case, according to property P2,
$\zeta_\beta \equiv \zeta$ does
not depend on the spatial position of galaxies and therefore
on any function $u_\beta({\bf r})$.

It thus turns out that any statistical test of independence between
$\zeta_\beta$ and
$u_\beta$
provides us with an unbiased estimate of the value of $\beta^{\star}$.
In particular the coefficient of correlation
$\rho(\zeta_\beta,u_\beta)$ has to vanish when
$\beta  = {\beta^{\star}}$, i.e.
\begin{equation}\label{Correlationcoefficient}
\beta  = {\beta^{\star}}
\,\,\,\,\,
\Longleftrightarrow
\,\,\,\,\,
\rho(\zeta_\beta,u_\beta)=0
\end{equation}
As revealed by Eq. (\ref{Correlation1}), the accuracy of this
estimator is related to the amplitude of the correlation between
$\left (\ln f \right )^\prime (M_\beta)$ and
$\zeta_\beta$. The steeper the function 
$\left (\ln f \right )^\prime$, or in other words the smaller the
dispersion of the luminosity function $f(M)$, the more accurate
is the estimate of the velocity parameter $\beta$, as expected.
In practice, this
accuracy can be obtained through
numerical simulations by analysing the influence
of sampling fluctuations on the coefficient of correlation
$\rho(\zeta_\beta,u_\beta)$.

The presence of a small-scale velocity dispersion (say of amplitude 
$\sigma_v$), not described
by the velocity model $v_\beta$, introduces according to Eq. (\ref{mubetabis})
a correlation between the derived quantities $\mu_\beta$ and $M_\beta$,
and consequently between the variables 
$\mu_\beta$ and $\zeta_\beta$.
Anyway since it is the correlation between the velocity model
$u_\beta$ and $\zeta_\beta$ which is considered herein, and because 
the random velocity noise is not supposed to be 
correlated with  
$u_\beta$,  the presence of a small-scale velocity dispersion
is not expected to drastically bias the estimator proposed Eq. 
({\ref{Correlationcoefficient}), at least as long as the 
variations of the quantity $u_\beta({\bf r})$ are smooth 
at the scale $\sigma_v$. 

Thanks to the introduction of the random variable $\zeta$, an unbiased
estimate of the parameter $\beta$ has indeed been obtained 
using a null-correlation technique. Null-correlation approaches
are characterised, in general, by their robustness --
i.e. some of the functions entering the statistical model are not required
to be fully specified (see for example Fliche \& Souriau 1979, 
Bigot et al. 1991, Triay et al. 1994, Rauzy 1997). 
Unlike the maximum likelihood methods, e.g.
the method proposed by Choloniewski (1995) and      
the VELMOD method of Willick et al. (1997a), no a priori
assumptions have been made here concerning the specific
shape of the luminosity function and
the spatial distribution of
the sources. In particular homogeneous as well as inhomogeneous
Malmquist biases are automatically accounted for.
Note also that selection effects in distance or redshift are 
allowed since Eq. (\ref{dP_4}) accepts any extra terms of the form
$\psi(\mu,u_\beta)$.

\subsubsection{Orthonormal decomposition of the velocity field}

It is worthwhile to mention that, with
regard to Eq. (\ref{Correlation1}), the orthonormal decomposition of
the velocity field, proposed in the ITF method
(Nusser \& Davis 1995), may be also applied herein. To see this we proceed
as follows.

It is assumed hereafter that there exists a N-dimensional vector 
${\bf \beta^{\star}}=(\beta_1^{\star},
\beta_2^{\star},...,\beta_N^{\star})$ such that
the quantity  $u({\bf r})$ can be decomposed as 
\begin{equation}\label{umodel1}
u({\bf r}) \equiv u_{\bf \beta^\star}({\bf r}) =
\sum_{j=1}^N \beta_j^{\star} \,u_j({\bf r})
\end{equation}
where 
$u_1({\bf r})$, $u_2({\bf r})$, ..., $u_N({\bf r})$ is a
set of functions verifying the following orthonormality condition,
\begin{equation}\label{orthonormality}
{\rm Cov}(u_i,u_j) = \delta_{ij}
\end{equation}
with 
$\delta_{ij}$ the Kronecker symbol and 
${\rm Cov}(u_i,u_j)$ the covariance of $u_i$ and $u_j$ on the sample.
Note that within the approximation 
$\beta_j^\star u_j({\bf r}) \ll 1$ (or equivalently
$\beta_j^\star v_j({\bf r}) \ll cz$),
Eq. (\ref{ubeta}) implies that
\begin{equation}
 v({\bf r}) \simeq 
\sum_{j=1}^N \beta_j^{\star} \,v_j({\bf r})
\end{equation}
The coefficient $C_\beta$ introduced in Eq. 
(\ref{Correlation1}) rewrites, as long as 
$(\beta_j-\beta_j^\star) u_j({\bf r}) \ll 1$ is satisfied, as 
\begin{equation}\label{Correlation2}
C_\beta 
\simeq 1 +
\,\left (\ln f \right )^\prime (M_\beta)
\times
\sum_{j=1}^{N} (\beta_j
-{\beta_j^{\star}})\,u_j({\bf r})
\end{equation}
which implies that for each function $u_i({\bf r})$,
\begin{equation}\label{Correlation3}
\rho(u_i,\zeta_\beta) \propto \sum_{j=1}^{N} (\beta_j
-{\beta_j^{\star}})\,{\rm Cov}(u_i,u_j)
\end{equation}
and thus, because of the orthonormality condition 
of Eq. ({\ref{orthonormality}), that
\begin{equation}\label{Correlationcoefficient2}
\beta_i  = {\beta_i^{\star}}
\,\,\,\,\,
\Longleftrightarrow
\,\,\,\,\,
\rho(\zeta_\beta,u_i)=0
\end{equation}
which provides us with the statistically independent estimates of the
$N$ parameters
$\beta_1^{\star},
\beta_2^{\star},...,\beta_N^{\star}$. The procedure for
constructing the orthonormal family 
$u_1({\bf r})$, $u_2({\bf r})$, ..., $u_N({\bf r})$ from
an arbitrary set of $N$ independent functions is described
in 
Nusser \& Davis (1995).

\section{Application to the IRAS sample}

\begin{figure}
\vbox
{\epsfig{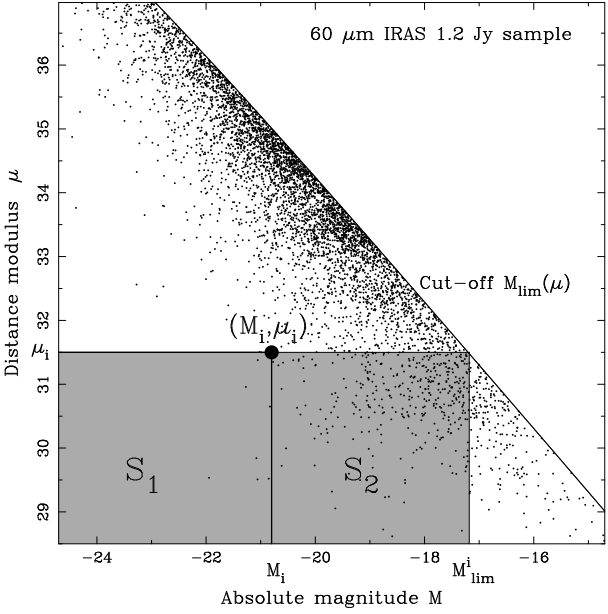}}
\caption{
 Distance modulus versus absolute magnitude for the 60 $\mu$m
IRAS 1.2 Jy sample (5321 galaxies). The procedure for estimating
the random variable $\zeta$ from
Eq. (6) is illustrated.
}
\label{figure1}
\end{figure}

The method described above is herein applied to the 
60 $\mu$m IRAS 1.2 Jy sample
(Fisher et al. 1995), a magnitude-redshift catalogue containing
$5321$ galaxies and complete up to a flux $F_{60}=1.2$ Jy.
Distance modulus and absolute magnitude are computed
using $H_0=100$ km s$^{-1}$ Mpc$^{-1}$ for the value of the Hubble
constant
(the cut-off in apparent magnitude is expressed as $m_{\rm lim}=14.3187$
with this notation). The magnitudes have been k-corrected
assuming a spectral slope $\alpha=-2$, implying that the
maximum absolute magnitude
introduced in Eq. (\ref{zeta}) reads as
$M_{\rm lim}(\mu)=m_{\rm lim} - {\rm k_{cor}}(\mu) - \mu$.
The distribution of the sources in the $M$-$\mu$ plane is
shown figure 1.

The peculiar velocity field model tested is the
one-parameter predicted IRAS velocity field
characterised by the parameter $\beta=\Omega_0^{0.6}/b_I$
(Strauss et al. 1992).

\begin{figure}
\begin{center}
\vbox
{\epsfig{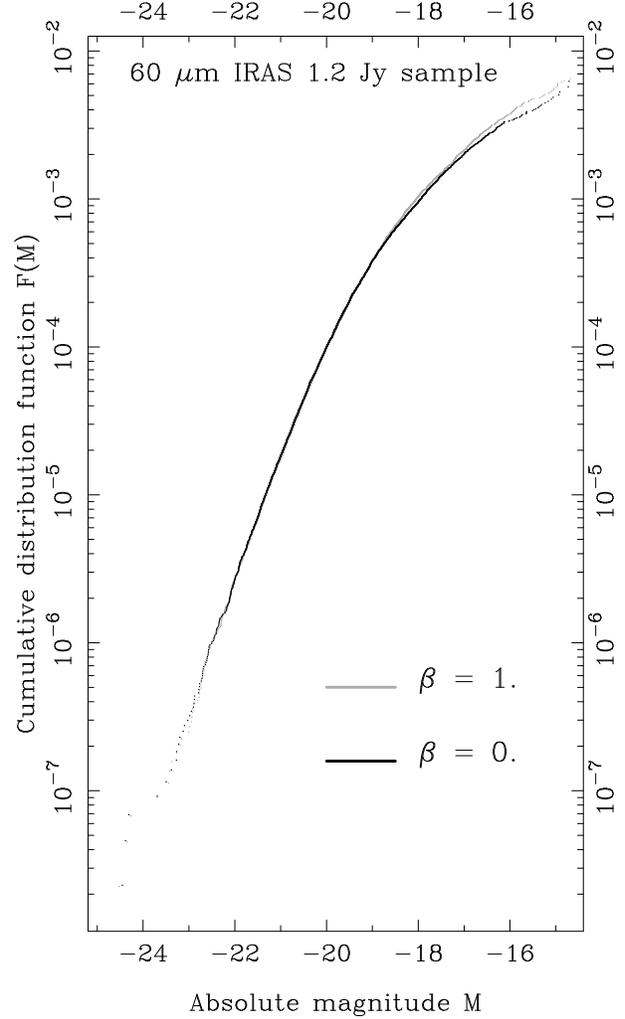}}
\caption{
Reconstruction of the Cumulative Luminosity Function of the
IRAS 1.2 Jy sources using the C$^-$ method of Lynden-Bell (1971).
The reconstructed CLF depends on the assumed velocity model.
}
\end{center}
\label{figure2}
\end{figure}

\subsection{Using apparent magnitude as the distance indicator}

We have first applied the method using the apparent magnitude
of IRAS galaxies as the distance indicator (restricting the analysis
to the 4115 galaxies within the redshift range
1000-12000 km s$^{-1}$). The Cumulative Luminosity Function (CLF)
$F(M)$ is presented in
figure 2. Here the CLF has been reconstructed using
the $C^{-}$ method of Lynden-Bell (1971) for two different velocity
models, $\beta=0$ corresponding to the case where peculiar velocities
are neglected and $\beta=1$. Note that 
the presence of peculiar velocities does not affect drastically
the CLF reconstruction.
The luminosity function of the IRAS galaxies
does not exhibit any turnover towards the
faint-end tail, at least within the observed range of magnitudes.
This means that, because of the large dispersion of such a distance 
indicator and even if the number of galaxies is large, one
cannot expect very strong constraints on the velocity model tested.

The correlation between the random variable $\zeta_\beta$ and
the velocity modulus $u_\beta$ for $\beta=0.6$ is shown in figure 3.
Variations of the coefficient of correlation as a function of
the parameter $\beta$ are given in figure 4. This curve is a monotonic
function, as expected.   
The preferred value of $\beta$ is the one corresponding to
$\rho(\zeta_\beta,u_{\beta})=0$ (here $\beta=0.1-0.15$).

Monte Carlo simulations
have been used to calculate the discrepancy between
$\rho(\zeta_\beta,u_{\beta})$
and zero due to sampling fluctuations. 
Each simulation is a sample containing 
$N_{\rm gal}=4115$ galaxies with the same $\mu_\beta$ and 
$u_\beta$  as observed and for which the random variable $\zeta_\beta$
is computed following Eq. (\ref{zetaestimate}) where the rank $r_i$ is
randomly generated according to a discrete uniform distribution 
between $1$ and $n_i$.  
The cumulative distribution function
of $\rho(\zeta_\beta,u_{\beta})$ obtained from a large number of simulations,
under the null hypothesis that the true value of 
$\rho(\zeta_\beta,u_{\beta}) = 0$, 
is shown in figure 5. Note that in practice this distribution does not
depend on the amplitude of the quantity $u_\beta$ since the coefficient
of correlation is a scale-free estimator of the correlation between two
random variables.  

The cumulative distribution function allows us to evaluate
the probability that the observed $\rho$ is by chance
greater than a given value, due to sampling fluctuations. 
A one-sided rejection test for the $\beta$ parameter
can thus be constructed. For example, the models
with $\beta \ge 0.7$ can be rejected with a confidence level of
95\% and $\beta \ge 1.1$ with a confidence level of
99\%. So our method applied to the IRAS sample, using the apparent magnitude
of the galaxies as a distance indicator, permits already at this stage
to reject high values for the parameter $\beta$.

\begin{figure}
\vbox
{\epsfig{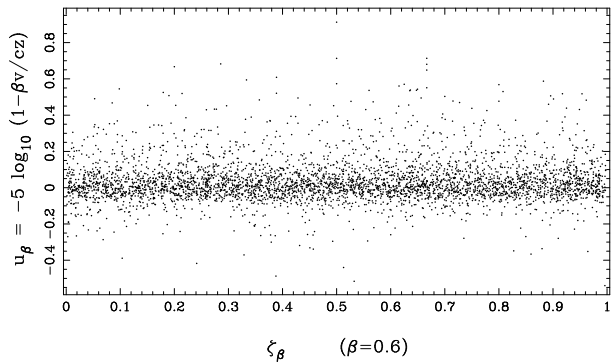}}
\caption{
Scatterplot indicating the correlation between the velocity modulus 
$u_\beta$ and the random variable $\zeta_\beta$ for $\beta=0.6$.
A observed sample correlation coefficient of 
$\rho(\zeta_\beta,u_{\beta}) \simeq 0.02$ was found in this case.}
\label{figure3}
\end{figure}

\begin{figure}
\vbox
{\epsfig{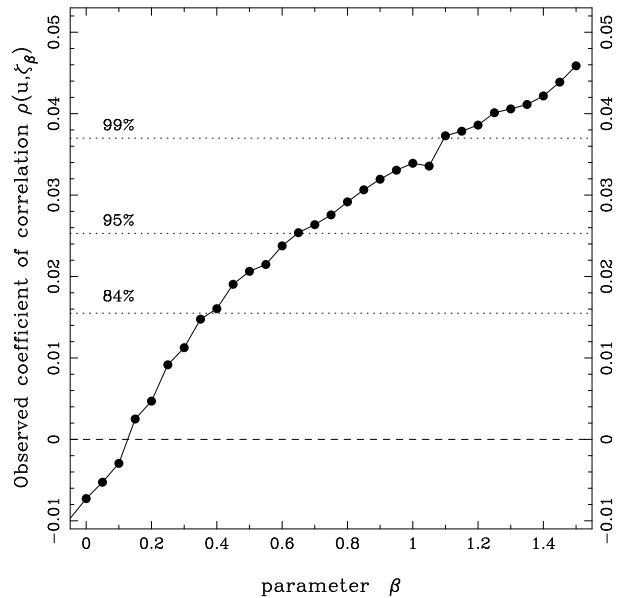}}
\caption{
The observed coefficient of correlation
$\rho(\zeta_\beta,u_{\beta})$ as a
function of the
parameter $\beta$. Confidence levels of rejection for
the parameter $\beta$ are calculated from figure 5. 
}
\label{figure4}
\end{figure}

\begin{figure}
\vbox
{\epsfig{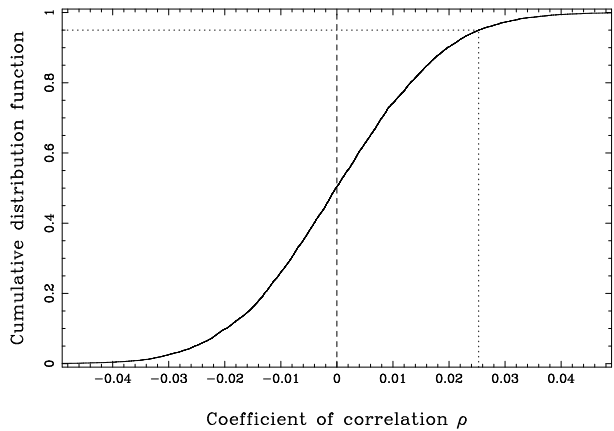}}
\caption{
Cumulative distribution function of $\rho$ due to sampling fluctuations.
This curve has been obtained by Monte Carlo simulations (see text).
}
\label{figure5}
\end{figure}

\begin{figure}
\vbox
{\epsfig{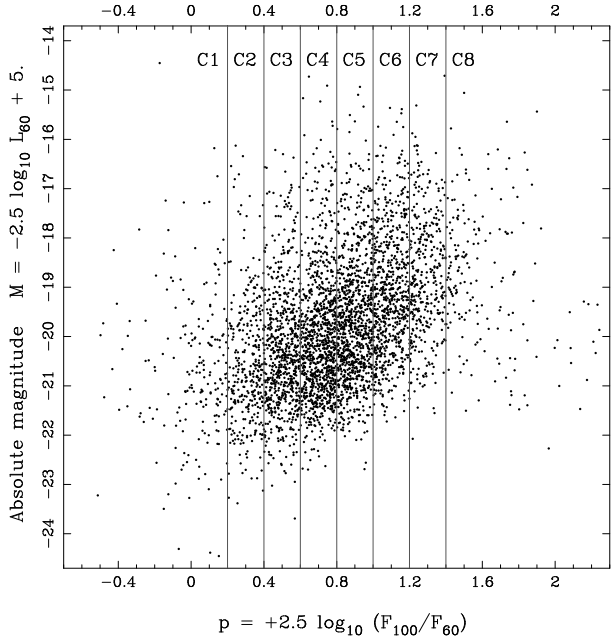}}
\caption{
 Observed correlation between the ``colour index" 
$p=2.5 \log_{10} (F_{100}/F_{60})$
and the absolute
magnitude $M$. The sample has been subdivided into $8$ classes,
$C_i$ according to the value of $p$. 
}
\label{figure6}
\end{figure}

\begin{figure}
\begin{center}
\vbox
{\epsfig{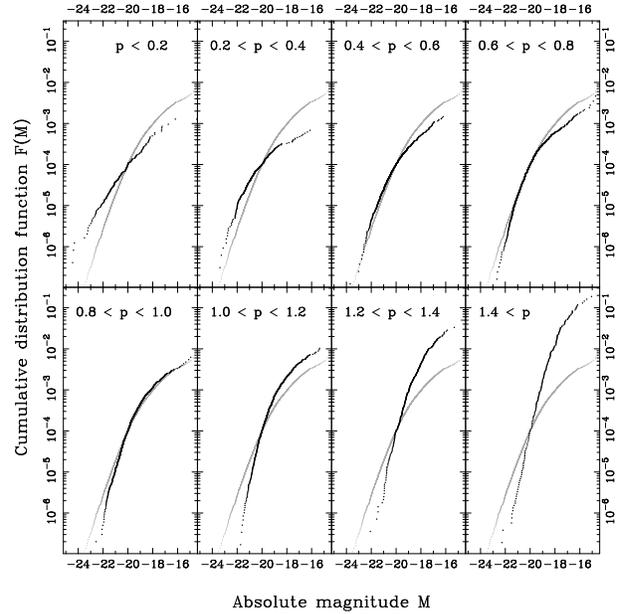}}
\caption{
C$^{-}$ reconstruction of the Cumulative Luminosity Function for the
$8$ individual $C_i$ classes in $p$ (dark curves). The global CLF
(i.e. no class) of figure 2 is shown for comparison (grey curve). 
The CLF's have been 
arbitrarily normalised to $10^{-4}$ at $M=-20.$
}
\end{center}
\label{figure7}
\end{figure}

\begin{figure}
\begin{center}
\vbox
{\epsfig{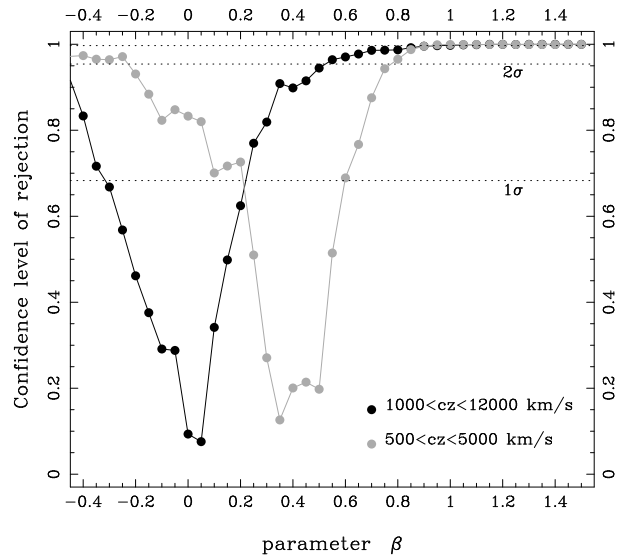}}
\caption{
Confidence level of rejection for the parameter $\beta$ (see text).
}
\end{center}
\label{figure8}
\end{figure}

\begin{figure}
\begin{center}
\vbox
{\epsfig{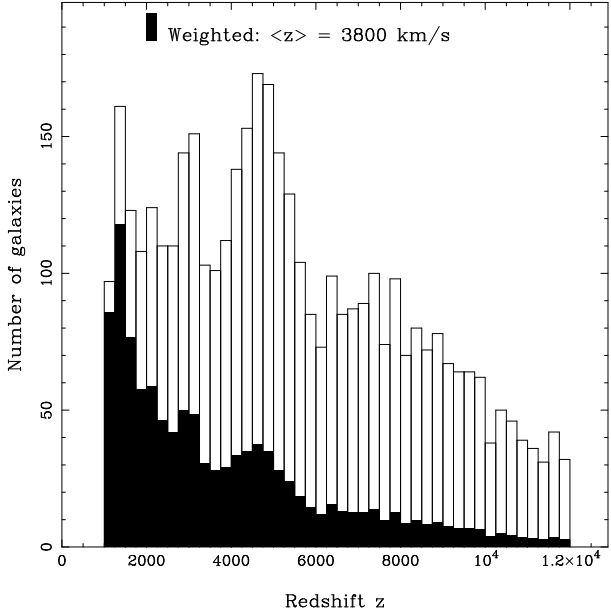}}
\caption{
Redshift distribution of the IRAS galaxies with 
$z \in [1000,12000]$ km s$^{-1}$ (4115 galaxies).
The dark histogram represents this distribution when
accounting for the natural weight  $\propto 1/z$. 
}
\end{center}
\label{figure9}
\end{figure}

\begin{figure}
\begin{center}
\vbox
{\epsfig{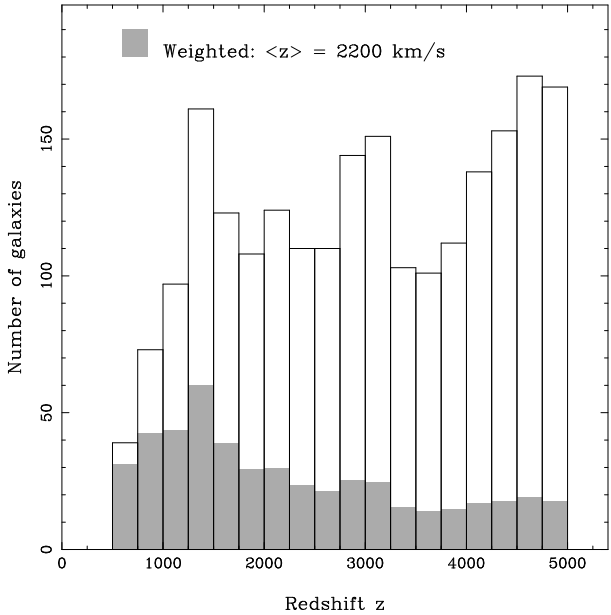}}
\caption{
Redshift distribution of the IRAS galaxies with 
$z \in [500,5000]$ km s$^{-1}$ and galactic latitude
$|b| > 20$ (1621 galaxies).
The grey histogram 
accounts for the natural weight  $\propto 1/z$. 
}
\end{center}
\label{figure10}
\end{figure}

\subsection{Introduction of a second parameter}

In a second step, the analysis is refined by taking into account 
the observed correlation between
the absolute magnitude $M$ and some ``colour index" defined
as $p=2.5 \log_{10} (F_{100}/F_{60})$ (with $F_{100}$ the
flux at 100 $\mu$m). The data
have been grouped in $8$ classes by interval of $p$
(see figure 6). Because of the (weak) correlation between $p$ and
$M$, the spread of the luminosity function
for each of these classes taken individually is expected to be
slightly smaller than the spread of the global luminosity function,
and thus the accuracy of the distance indicator somewhat improved.
Figure 7 illustrates such a trend.

For each individual class, the random variable $\zeta_\beta$ is computed
according to Eq. (\ref{zetaestimate}).
The correlation between $\zeta_\beta$ and
the velocity modulus $u_\beta$ is after that evaluated for the whole sample.
The influence of sampling fluctuations is estimated from Monte Carlo 
simulations, as described in the previous section.
The results are presented in figure 8 in terms of the confidence level
of rejection for the parameter $\beta$. More explicitly, the
quantity plotted in ordinate is $1-2  \,
{\rm Prob}(\rho \le -|\rho_{\rm obs}(\beta)|)$
where the probability that  
the coefficient of correlation $\rho$ is less than 
$-|\rho_{\rm obs}(\beta)|$ due to sampling fluctuations
is given by the cumulative distribution function
of $\rho$.

The method was first applied to the galaxies in the redshift range
1000-12000 km s$^{-1}$.  It is found that $\beta \in [-0.35,0.25]$ at
$1\sigma$, and that models with
$\beta \ge 0.5$ can be rejected with a confidence level of
95\%. This result is in disagreement with most of the
analyses based on Tully-Fisher data e.g. VELMOD on MarkIII
(Willick \& Strauss 1998), ITF method on SFI (Da Costa et al. 1998),
ROBUST method on MarkIII MAT sample (see next section),
favouring a value of $\beta \simeq 0.5$.
We interpret this discrepancy as follows.

When fitting a velocity model to data, the natural weight assigned
by the fitting procedure
to each galaxy is roughly proportional to the inverse of its redshift,
because the accuracy of the distance indicator decreases as $1/z$.
The mean effective depth of the volume where the velocity model is
compared to data has to be estimated using these weights. For
our first sample with $z \in [1000,12000]$ km s$^{-1}$,  we find
a mean effective depth
of $3800$ km s$^{-1}$ (see figure 9).

In order to mimic the effective volume sampled by Tully-Fisher data
we have applied the method to a truncated version of the IRAS sample
containing 1621 galaxies with $z \in [500,5000]$ and galactic latitude
$|b| > 20$ (the mean effective depth of this sub-sample is now
2200 km s$^{-1}$, see figure 10).
Figure 8 shows that the value of $\beta$ estimated from this
truncated sample is fully consistent with the values obtained using
Tully-Fisher data. An interpretation of these results could
be that the predicted IRAS velocity field model, while successful
in reproducing locally the cosmic flow, fails to describe the
kinematics on larger scales.
 
However, as pointed out by our anonymous referee, the results
derived above apply only to the extent that the
photometry of the IRAS sample does not suffer from systematic
errors. It is worthwhile to stress again that the philosophy of
the ROBUST method is to impose very few assumptions -- only that 
the luminosity function is independent of position, that the sample 
is strictly complete in apparent magnitude and that redshift and 
apparent magnitude measurements are not affected by systematic biases.
Thus, our analysis could be 
affected if the IRAS photometry were (mildly) non-uniform. 
To evaluate the amplitude of such effects on our results would essentially 
require to adopt a realistic model for these systematic photometry variations.
Since our primary goal here is to present a new method, we feel that such an 
error analysis is unwarranted. It is also worth noting that systematic errors 
in the IRAS photometry -- if indeed present -- would also affect methods for 
reconstructing the IRAS predicted peculiar velocity model. In particular, 
they would generate some systematic variations in the spatial selection 
function entering the weighting scheme used in density and velocity 
reconstruction (Strauss et al. 1992, Branchini et al. 1999), leading to
systematic discrepancies between the IRAS predicted velocity field and the 
true cosmic flow. Moreover, it is interesting to note that the ROBUST method 
could easily be extended to provide a non-parametric test of the
uniformity of photometry in redshift surveys; we will consider such an 
application in future work.

\begin{figure}
\begin{center}
\vbox
{\epsfig{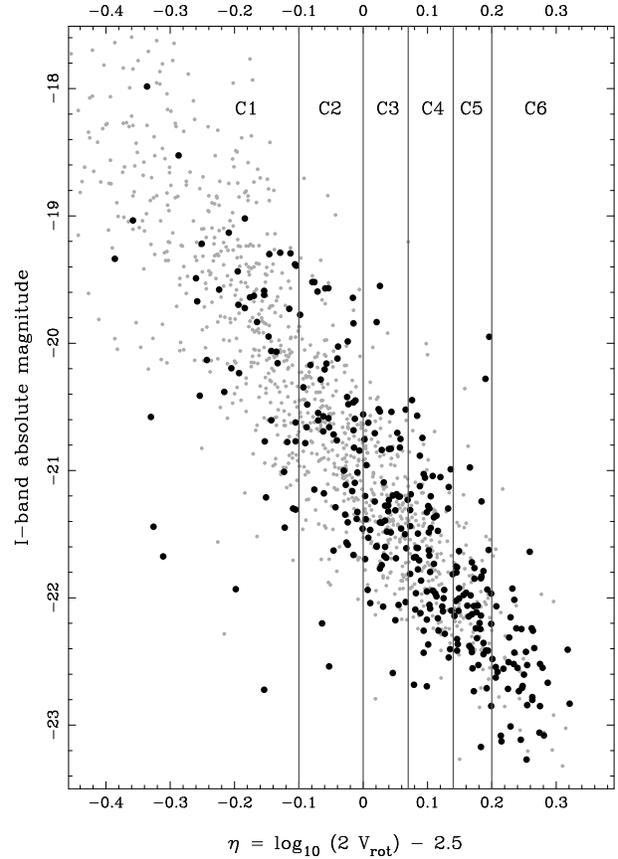}}
\caption{
The Tully-Fisher relation for the MarkIII MAT sample
(Willick et al. 1997b). A subsample of 318 galaxies (dark dots)
complete in apparent magnitude up to $m_{\rm lim}=11.25$ mag. has been
extracted. This subsample is divided in $6$ classes $C_i$ containing
approximately $50$ galaxies each.
}
\end{center}
\label{figure11}
\end{figure}

\begin{figure}
\begin{center}
\vbox
{\epsfig{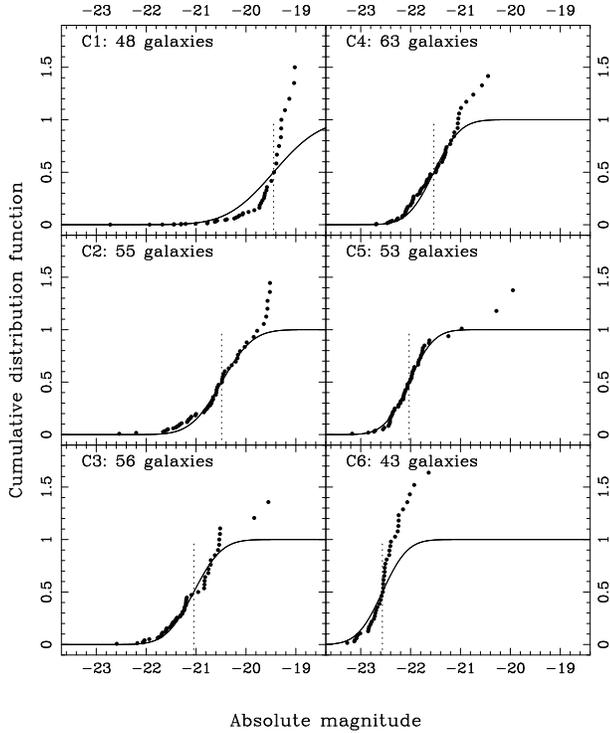}}
\caption{
For each of the $6$ classes $C_i$, the $C^{-}$ reconstructed
Cumulative Luminosity Function (dark dots). The solid curves
are the expected cumulative distribution function of $M$
assuming a linear Direct (i.e. Forward) Tully-Fisher relation
with Gaussian residuals and calibrated using the values proposed
in Willick et al. (1998) (see text). For convenience in comparison, the
$C^{-}$ reconstructed CLF's $F_{\rm rec}(M)$ have been normalised such that
$F_{\rm rec}(M_i)=0.5$ at $M_i$ the mean absolute magnitude for each
classes $C_i$ (dotted lines).
}
\end{center}
\label{figure12}
\end{figure}

\begin{figure}
\begin{center}
\vbox
{\epsfig{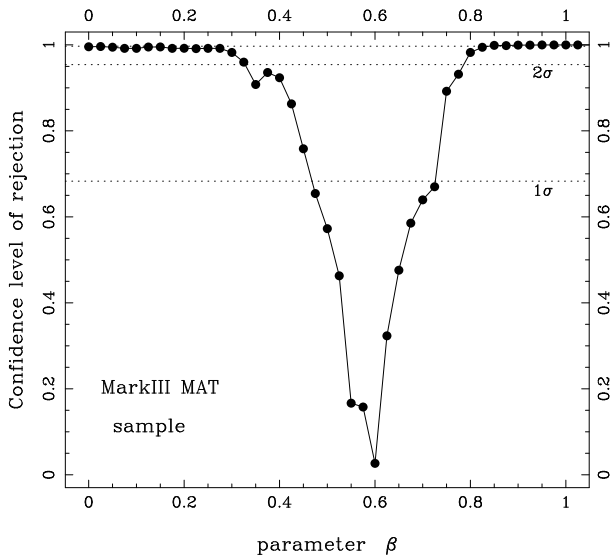}}
\caption{
Confidence level of rejection for the parameter $\beta$
($\beta = 0.6 \pm 0.125$).
}
\end{center}
\label{figure13}
\end{figure}

\begin{figure}
\begin{center}
\vbox
{\epsfig{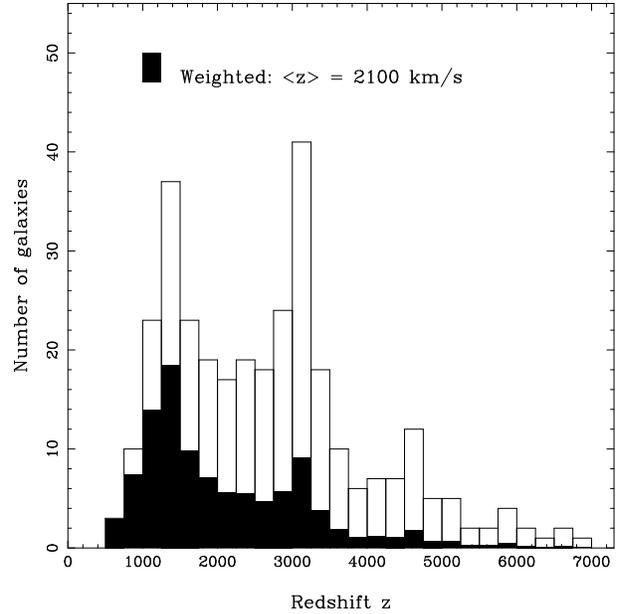}}
\caption{
Redshift distribution of the $318$ MarkIII MAT galaxies with
$z \ge 500$ km s$^{-1}$ and apparent I-band magnitude $m \le 11.25$
mag.
The dark histogram
takes into account the natural weight  $\propto 1/z$.
}
\end{center}
\label{figure14}
\end{figure}

\section{Application to the MarkIII MAT sample}

We have shown in the previous section that our method
permits to extract valuable information on the cosmic 
velocity field by using the fluxes of the IRAS galaxies
as a distance indicator. The potential of our method
is illustrated in this section by treating a more classical
case: the Tully-Fisher MarkIII MAT sample
(Willick et al. 1997b and references therein).

In a first step we have selected from the 1355 galaxies
of the MarkIII MAT catalog a subsample for which the
selection effects in apparent magnitude are well described by
a Heaviside cut-off, $\theta(m-m_{\rm lim})$. 
Assuming that the galaxies are homogeneously distributed
in space, the value of $m_{\rm lim}$ can be found
by analysing
the variations in the logarithm of the cumulative count as a
function of the limit in apparent magnitude -- see for
example figure 4 in Rauzy (1997). 
In a separate paper (Rauzy, in preparation) we propose
anyway a
simple tool for assessing the completeness in apparent magnitude
of redshift-magnitude catalogues which does not require any
assumption concerning the spatial distribution of the sources. This method
is closely based on the statistical test presented in
Efron \& Petrosian (1992).
We have applied this test to the MarkIII MAT sample, finding
that the completeness in I-band corrected apparent magnitude
is satisfied up to $m_{\rm lim} = 11.25$ magnitudes. 

Discarding, in addition, the galaxies with
$z \le 500$ km s$^{-1}$, we are left with a subsample containing
$318$ galaxies. Note that the selection effects in 
redshift which affect the MarkIII MAT sample (Willick et al. 1996)
are not a source of problems for our method.
The Tully-Fisher relation for the $318$ extracted  galaxies
is shown in figure 11.
  
In a second step, we divided the subsample in $6$ classes $C_i$
according to the value of $\eta$, from the slower rotators in $C_1$
to the faster rotators in $C_6$ (see figure 11).
The C$^-$ reconstructed cumulative 
luminosity function is shown for each class
in figure 12. If the binning in the log line-width parameter $\eta$ were
narrow enough, one would expect the reconstructed  luminosity function
for the class $C_i$ 
to be centered on $M_i = a_{\rm DTF}\, \eta_i + b_{\rm DTF}$ and of dispersion
$\sigma_{\rm DTF}(\eta_i)$,  where $\eta_i=<\eta>_{C_i}$ denotes the mean 
of the $\eta$'s in $C_i$
and  
$a_{\rm DTF}$, $b_{\rm DTF}$ and 
$\sigma_{\rm DTF}(\eta_i)$ are respectively the slope, zero-point
and dispersion of the Direct (i.e. Forward) Tully-Fisher relation.
We plot in figure 12 the expected culmulative luminosity function
when a linear DTF relation with gaussian residuals is assumed and
using the values proposed in Willick et al. (1998) for the
calibration parameters. The width of the bins has been accounted for
by adding in quadrature to the dispersion 
$\sigma_{\rm DTF}(\eta_i)$ the product of 
$|a_{\rm DTF}|$ and the standard deviation 
of the $\eta$'s in each class $C_i$ .
For convenience in comparison, Each C$^-$ reconstructed CLF
$F_{\rm rec}(M)$ has been normalised such that
$F_{\rm rec}(M_i)=0.5$. 

It is not clear from figure 12 whether
a linear DTF relation with gaussian residuals is plainly
successful in reproducing the data. Note in particular that
the C$^-$ reconstructed CLF's for the slow rotators do not
exhibit a turnover towards the faint magnitudes. To question
the validity of a linear Tully-Fisher relation is anyway beyond the scope
of the present paper. At this stage it is however worthwhile to mention
that our method makes a very conservative use of Tully-Fisher
information. In fact, we require to assume neither a linear TF relation
nor a gaussian distribution for its residuals.

The result of the ROBUST method applied to the MarkIII MAT catalog
is shown in figure 13. The analysis was performed as described
in section 3.2. The mean effective depth of the subsample is
2100 km s$^{-1}$ (see figure 14).
We find a value of $\beta=0.6 \pm 0.125$, in complete agreement with
the VELMOD and ITF method applied to similar Tully-Fisher data
(Willick et al. 1997a and 1998, Davis et al. 1996, Da Costa et al. 1998).

We want however to emphasise the robustness of our approach compared
to these two fitting methods. 
Firstly, no assumptions have been made herein concerning
the linearity of the Tully-Fisher law, as is required by both the VELMOD
and the ITF method. Secondly, we do not need the sample to be free of 
selection effects in the log line-width parameter $\eta$ as is the 
case for the ITF method. Thirdly, 
the spatial distribution of the sources, the selection
effects in redshift and the shape of the distribution function of the
TF residuals need not be specified, as is required by the 
maximum likelihood VELMOD method.
   
\section{Conclusion}

We presented a method for fitting peculiar velocity models
to complete flux limited magnitude-redshift catalogues, using
the luminosity function of the sources as a distance indicator, i.e.
assuming that the distribution function of the absolute magnitudes of
the galaxies does not depend on the spatial position. 

Our method is based on a null-correlation approach.
For a given peculiar velocity field model parametrised by 
a parameter $\beta$,
we defined a random variable $\zeta_\beta$, computable from the observed
redshifts and apparent magnitudes of the sampled galaxies, which
has the property of being statistically independent on the position
in space (and thus on the modelled radial peculiar velocities
themselves) if and only if the parameter $\beta$ matches its true value
$\beta^\star$.
Therefore any test of independence between the random variable
$\zeta_\beta$ and the modelled velocities or similar quantities
provides us with an unbiased estimate of the value of 
$\beta^\star$. The method can be easily generalised
to velocity models parametrised by an $N$-dimensional vector
${\bf \beta}=(\beta_1,\beta_2,...,\beta_N)$.
 
The method is characterised by its robustness.
No assumptions are made concerning
the spatial distribution of sources and their luminosity
function and selection effects in redshifts are also allowed.
The required strict completeness in apparent magnitude can moreover
be checked independently (Rauzy, in preparation). 
Furthermore the inclusion of additional observables correlated with
the absolute magnitude
is straightforward.   

The predicted IRAS peculiar velocity model characterised by the density
parameter $\beta$ 
has been tested on two samples, the Tully-Fisher MarkIII MAT sample and
the 60 $\mu$m IRAS 1.2 Jy sample using the fluxes as the distance indicator.

The application of our method to the MarkIII MAT sample gives 
a value of $\beta=0.6 \pm 0.125$, in excellent agreement with the results
obtained previously by the VELMOD and ITF methods on similar 
datasets. Our method is however more robust than these two fitting methods.
In particular, we make a very conservative use
of the Tully-Fisher information. We do not require to assume
the linearity of the Tully-Fisher relation nor a gaussian distribution
of its residuals.

We showed that our method allows to extract some valuable 
informations on the peculiar
velocity field from the fluxes  of
the IRAS 1.2 Jy sample. The poor accuracy of the distance indicator
(due to the broad spread of the luminosity function)
is balanced in this case thanks to the large number of galaxies contained in
the sample. The IRAS sample permits to probe the cosmic flow at larger
scales. Indeed, the mean effective depth of the volume 
in which the velocity
model is compared to the data is almost twice the mean effective depth of
the MarkIII MAT sample.  

The application of our method to 
an IRAS subsample truncated in distance, of an effective depth similar to
the MarkIII MAT sample, gives a value of $\beta$ in accord with
the values obtained using Tully-Fisher data. On the other hand
when the application is performed on the whole sample, we found
that the predicted IRAS velocity models with $\beta \ge 0.5$ can be
rejected with a confidence level of $95\%$.
These results suggest that the predicted IRAS velocity model,
while successful
in reproducing locally the cosmic flow, fails to describe the
kinematics on larger scales.  

Note that these results do not lead to dismiss
the linear ``biasing'' paradigm.
As the errors on the predicted IRAS velocity field
increase with distances, it could be that the predictions at the 
scales considered herein, i.e. 
beyond $5000$ km s$^{-1}$, drastically differ from the true cosmic flow
(see for example Davis et al. 1995).  


\section*{Acknowledgements}
We are thankful to Michael Strauss for
providing us with the predicted IRAS peculiar velocity model. SR acknowledges
the support of the PPARC and both authors acknowledge the use of the
STARLINK computer node at Glasgow University.

\end{document}